# Asymmetric frequency conversion with acoustic non-Hermitian space-time varying metamaterial

Xinhua Wen[1], Xinghong Zhu[1], Alvin Fan[1], Wing Yim Tam[1], Jie Zhu[2], Fabrice Lemoult[3], Mathias Fink[3,a)], Jensen Li[1,a)]

[1]Department of Physics, The Hong Kong University of Science and Technology, Clear Water Bay, Kowloon, Hong Kong, China

[2]Department of Mechanical Engineering, the Hong Kong Polytechnic University, Hung Hom, Kowloon, Hong Kong SAR, People's Republic of China

[3]Institut Langevin, ESPCI Paris, PSL University, CNRS, 1 rue Jussieu, 75005 Paris, France

**Abstract**

Space-time modulated metamaterials support extraordinary rich applications, such as parametric amplification, frequency conversion and non-reciprocal transmission. However, experimental realization of space-time modulation is highly non-trivial, hindering many interesting physics that are theoretically predicted to be experimentally demonstrated. Here, based on the proposed virtualized metamaterials with software-defined impulse response, we experimentally realize non-Hermitian space-time varying metamaterials for efficient and asymmetric frequency conversion by allowing material gain and loss to be tailor-made and balanced in the time domain. In the application of frequency conversion, the combination of space-time varying capability and non-Hermiticity allows us to diminish the main band through gain-loss balance and to increase the efficiency of side band conversion at the same time. In addition, our approach of software-defined metamaterials is flexible to realize the analogy of quantum interference in an acoustic system with design capability. Applying an additional modulation phase delay between different atoms allows to control such interference to get asymmetric amplification in frequency conversion.

In the past two decades, metamaterials with spatial modulation (with spatially varying parameters) have opened an emerging paradigm to manipulate classical waves, and have been used to achieve many intriguing physical phenomena like negative refraction and invisibility cloaking [1-6]. To date, the properties of most metamaterials are often locked into place once fabricated. To make further usage of metamaterials in practical scenarios, numerous efforts have been made to construct tunable or reconfigurable metamaterials, such as using coding or digital metamaterials [7,8], electronically-controlled active metamaterials [9-11] to achieve reconfigurable lensing, beam-steering, and tunable topological edge states [12,13].

Compared to metamaterials with the material profile only varying in the spatial domain, time modulation [14-18] greatly extends the degree of wave manipulation, and open completely new possibilities for novel physics like time reversal mirror [19,20], parametric amplification [21,22], frequency conversion [17] and non-reciprocal transport and transmission [14,15]. Interestingly, analogy concepts originally defined for spatially inhomogeneous metamaterial can be explored in the temporal domain now, including the concept of temporal effective media and time-gradient metasurfaces [23, 24]. In the past few years, we start to witness some initial experiments in time-varying metamaterials to construct time-varying gradient/Huygen's metasurfaces in microwaves, and acoustic cavities with space-time modulation [25-27], proving time-varying metamaterials are indeed plausible. However, achieving a time-varying metamaterial with time-varying material parameters with arbitrary specifications and high working efficiency are both nontrivial. For example, adding non-Hermiticity, or gain-loss contrast, into a time-varying system allows efficient non-reciprocal frequency conversion in a recent theoretical proposal [28].

In this work, we focus on using virtualized active metamaterials to realize time-varying capability. Such an approach has already allowed decoupled tunable resonating frequencies, strength and linewidth [11], which can now be further made time-varying simply through a computer program. More importantly, the virtualized approach allows us to have both active and passive media with controllable gain and loss to construct a time-varying non-Hermitian system, which is the central focus in this work. We experimentally demonstrate that efficient asymmetric frequency conversion is possible with amplification and main-band elimination at the same time due to the time-varying non-Hermiticity. Moreover, due to the flexibility of our software-defined metamaterial platform, we can arbitrarily define the "energy"-level diagrams to manipulate the frequency conversion, in analogy to quantum interference effect, but with much more flexibility in design than in the case of quantum materials.

In acoustics, one way to achieve space-time modulation is through tuning mechanical resonances [27] but the modulation frequency and depth cannot be easily altered greatly. Here we use a digitally virtualized active metamaterial approach [11] to experimentally realize the space-time modulation. The active metamaterials with software-defined frequency dispersion can adjust material gain and loss flexibly, which give us a versatile platform to study the interesting physics in the domain of non-Hermitian space-time varying metamaterials, with very flexible specifications of the resonance and modulation parameters.

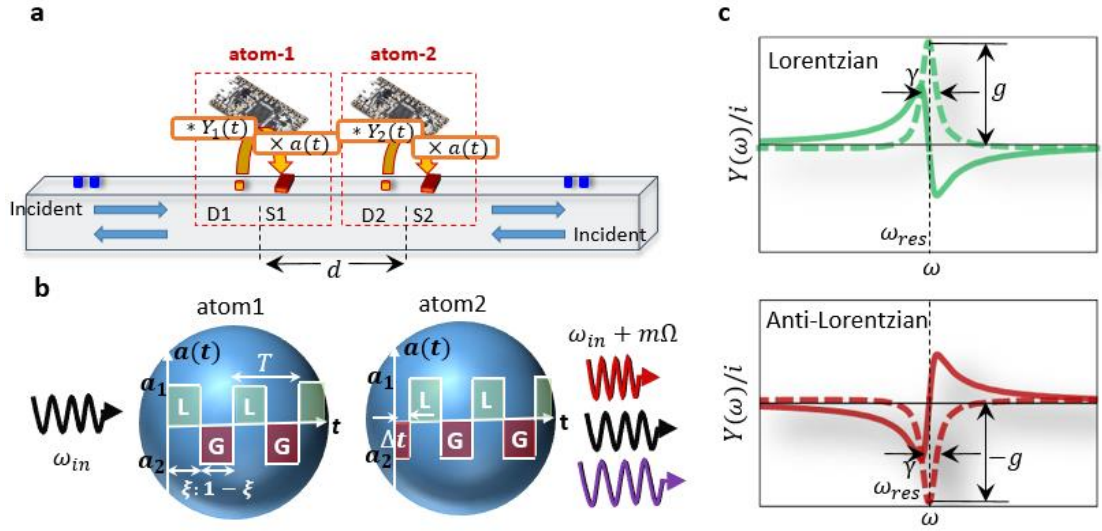

**Fig. 1 | Non-Hermitian space-time varying metamaterial platform. a**, The schematic diagram of the space-time varying metamaterials platform in a 1D acoustic waveguide. **b**, A square-wave amplitude modulation $a(t)$ is applied on the impulse response function of the meta-atoms with a modulation period $T$ (modulation frequency $F = 1/T$) and duty-cycle $\xi$. Moreover, an additional modulation time delay $\Delta t$ can be applied on atom2. We also define $\Delta\phi = 2\pi F\Delta t$ as the modulation phase delay. Consequently, frequency conversion occurs in such a space-time varying metamaterial. **c**, The convolution function $Y(f)/i$ for a passive atom (upper panel) and active atom (lower panel), corresponds to the Lorentzian and anti-Lorentzian line-shape respectively. Real and imaginary parts are plotted in solid and dashed lines respectively.

Figure 1**a** schematically shows our time-varying metamaterial platform in a 1D acoustic waveguide. It consists of two digitally virtualized metamaterial atoms, each represented by one microphone ($D_i$) and one speaker ($S_i$) with a feedback between them through an external microcontroller to perform a digital convolution $Y_i$. The virtualized meta-atom with mathematically designed convolution kernel $Y_i$ gives us the flexibility to realize a Lorentzian or anti-Lorentzian resonance, schematically shown in Fig. 1(c)

for a passive atom (upper panel) and active atom (lower panel) respectively. The passive and active atoms with the same resonating frequency $f_{res}$, resonating linewidth $\gamma$ but opposite resonating strength $g$ produce the balanced material loss and gain, which is useful for us to exploit the interplay between gain and loss in a non-Hermitian time-varying system.

For implementation of static atoms, the signal arriving at the microphone $D_i$ is detected, digitally sampled and convoluted with the convolution function $Y_i$, and then feed-backed to the speaker $S_i$ to generate scattered waves. To obtain a time-varying atom, we apply an amplitude modulation, $a(t)$, on the convoluted signal before firing speaker $S_i$, e.g. at one of the atoms (atom1, Fig. 1**b**). As a typical example, a square-wave modulation $a(t)$ ($a(t) = a_1$ for $mod(t,T) < \xi$, otherwise $a(t) = a_2$) is applied on the atom1 with modulation period $T$ and dutycycle $\xi$. For simplicity, we use $a_1 = 1$ to represent the fore-mentioned passive atom and $a_2 = -1$ to represent the same atom with opposite resonating strength, i.e. an atom with material gain balanced with material loss of the passive atom. Consequently, the atom with such a time-varying response behaves as material loss and gain alternating in time. Particularly, by setting the duty-cycle $\xi = 0.5$, we can mimic a "parity-time" symmetric metamaterial with loss and gain balanced in the temporal instead of the spatial domain. Such a time-modulation can also be applied on another atom (atom2) to realize a space-time modulation with an additional time delay $\Delta t$ in the modulation, which induces modulation phase delay $\Delta \phi = 2\pi \Delta t/T$ on atom2. The space-time modulation, in general, induces frequency conversion in the scattered waves from an incident frequency $f_{in}$ to other output frequencies $f_{out}$ with a relationship $f_{out} = f_{in} + mF$, where $F = 1/T$ is the modulation frequency, integer $m$ is the order of the discrete Floquet modes. A positive (negative) $m$ represents an upward (downward) frequency conversion. In the following, we will investigate the cooperative effect between time-varying resonating parameters and gain-loss balance in such a process.

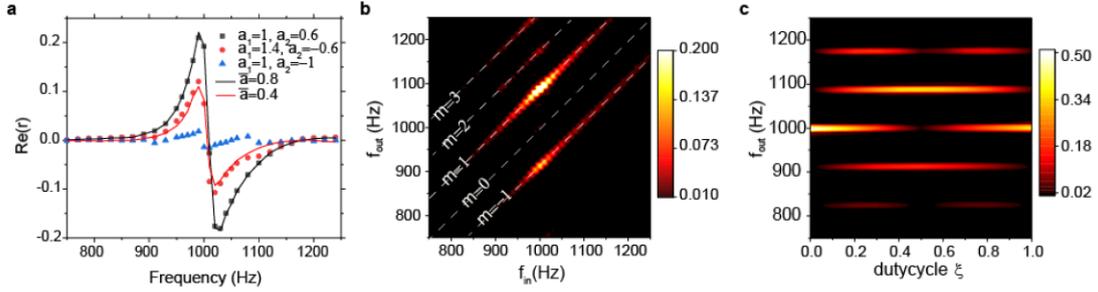

**Fig. 2|Efficient frequency conversion with temporal gain-loss balance at a single time-varying meta-atom. a**, The experimentally extracted real part of reflection coefficient for a time-varying atom with different modulation amplitudes (symbols) or the static counterpart with effectively averaged resonance strength (lines). **b**, The 2D map of the reflection amplitude (varying input and output frequencies $f_{in}$ and $f_{out}$) for a time-varying atom with balanced material gain and loss in time domain by setting the duty-cycle as 50%, clearly shows the suppressed 0-order due to the gain-loss balance in time domain and frequency conversion to generate discrete side-bands. **c**, The amplitude of reflection coefficient for different Floquet modes against the modulation duty-cycle $\xi$ at a fixed input frequency $f_{in} = 1000$Hz.

For a better illustration, we start with a single meta-atom, being static with a fixed convolution kernel of frequency spectrum $Y(f)$ (schematically shown in the upper panel of Fig. 1**c**)):

$$Y(f) = \frac{gf_{res}}{f_{res}^2 - (f+i\gamma)^2} e^{i2\pi f \delta t} \quad (1)$$

with $f_{res} = 1000$ Hz, $\gamma = 12$ Hz and $g = 14$ Hz. $\delta t = 160\mu s$ is the additional

program time-delay to compensate the electronic time delay of the microcontroller in order to make the reflection amplitude spectrum in Lorentzian shape. After applying an amplitude modulation, the resonance strength becomes $g \times a(t)$. For an atom with square-wave amplitude modulation, the resonance strength varies with time, i.e. becoming $g \times a_1$ for $mod(t,T) < \xi$, otherwise $g \times a_2$. Specifically, setting $a(t) = a_1 = a_2 = 1$ returns to the case of a static atom without time modulation.

Figure 2**a** shows the experimental results of the real part of reflection spectrum along the 1D waveguide for three amplitude modulation cases (symbols) with fixed duty-cycle $\xi = 0.5$ and modulation frequency $F = 90Hz$ and two static cases (lines). Applying a square-wave amplitude modulation with $a_1 = 1$ and $a_2 = 0.6$ on the atom to demonstrate the time-varying capability, we then obtain the reflection spectrum by firing an incident wave from one end of the waveguide at individual frequencies ($f_{in}$) and measure the reflected signal at the same frequency ($f_{out} = f_{in}$). The obtained reflection spectrum is plotted as symbols in black color. The real part of reflection spectrum of the time-varying atom shows a Lorentzian line shape as if it is a static atom. We found that if a static atom (black solid line) is chosen to have an effective modulation amplitude $\bar{a} = (a_1 + a_2)/2 = 0.8$, it matches well with the resultant spectrum of the time-varying atom. Therefore, the scattering property of the time-varying atom for the 0-order Floquet mode ($f_{out} = f_{in}$) is equivalent to that of a static atom with an averaged resonance strength. For our system that can go active, even if we apply a square wave modulation with negative $a_2$ which corresponds to the atom with material gain, the picture about temporally averaged resonance strength still holds.

For a time-varying atom with $a_1 = 1.4$ and $a_2 = -0.6$, the experimental real part of reflection spectrum (red symbols) also agrees well with that of a static atom (red solid line) with an effective modulation bias $\bar{a} = 0.4$. Such an averaged response is also found similarly in "temporal effective medium" where the modulation frequency is much higher than the signal frequency [23] except that here we are working in the opposite regime, in which the signal frequency (~1kHz) is much higher than the modulation frequency (90Hz). By using such an averaged resonance strength concept, we can actually obtain a nearly zero reflection amplitude by instructing balanced amplitude modulation ($\bar{a} = 0$), as shown in Fig. 2**a** using blue symbols (a square wave modulation with $a_1 = 1$ and $a_2 = -1$). Tailoring the impulse response in time domain provides a mechanism to create a medium with effectively averaged constitutive parameters for the zero-th order Floquet band. More information on the effective medium parameters can be found in Supplementary Information.

To get a full picture of the reflection properties for such a time-varying atom, we scan the input frequency ($f_{in}$ from 750 to 1250Hz with a step of 10Hz) to get reflection amplitude spectrum with the same output frequencies range ($f_{out}$, 750 to 1250Hz), and plot in Fig. 2**b** as a 2D color map. It is found that along the diagonal line $f_{out} = f_{in}$ (along the dashed line labeled as $m = 0$ band), the amplitude always stays low as expected due to the effective modulation bias $\bar{a} = 0$. In addition, frequency conversion occurs to generate the side bands $f_{out} = f_{in} + mF$ with integer $m$. From the Fig. 2**b**, it is found that the bands of $m = \pm 1$ are more prominent with significant amplitudes than other orders. We also note that the efficiency of the frequency conversion is much higher near the resonating frequency for $f_{in}$. In Fig. 2**b**, all the side-bands show a prominent amplitude peak around 1kHz, the chosen resonating frequency of the designed meta-atom. Moreover, the amplitude of up-converted Floquet mode is larger than that of the down-converted Floquet mode, this asymmetry results from the frequency dispersive response of the speakers (see supplementary materials for detail information).

Here, the frequency conversion can be understood as a cascade of individual transfer processes from a speaker $S_j$ to another speaker $S_i$ (can also be the same one) in the system: $S_j e^{i2\pi f_{in}\delta t_{ij}} Y_i(f_{in}) a_i(f_{out} - f_{in}) \to S_i$. When sound wave at frequency $f_{in}$ is emitted from speaker $S_j$, it propagates to the location of $D_i$ with a time delay indicated by $\delta t_{ij}$. The signal hits the resonating convolution kernel $Y_i(f_{in})$ and then is amplitude-modulated through $a_i$ (as a convolution in frequency domain) before firing again at speaker $S_i$. Such a process goes over between any pairs of speakers and happens with multiple times and with feedback. As a whole, the frequency conversion is enabled by $a(f_{out} - f_{in})$, which is completely dictated by the Fourier series of $a(t)$. The frequency conversion from $f_{in} \to f_{in} + mF$ is therefore corresponding to the Fourier component of $a(mF)$. With loss and gain balanced in our system, the Fourier series of $a(t)$ misses the constant term ($\bar{a} = 0$), i.e. the $m = 0$ band (the main band) diminishes. Moreover, the transfer process also explains why the frequency conversion is more efficient when the input frequency (instead of the output one) is matching to the resonating frequency through the term $Y_i(f_{in})$. With the tunable parameters (the modulation frequency, depth and the resonance) can be individually specified in a program within the microcontroller, the efficiency of frequency conversion can be adjusted flexibly and we will see that even an efficiency larger than 100% (Fig. 3**d** demonstrates such an example) is possible. Such an efficient operation is assisted by the gain in the system (i.e. power injected from external source).

Temporally balanced gain and loss in the system enhances the efficiency of frequency conversion for the time-varying atom, as gain-loss balance diminishes the main-band and greatly convert the energy to the side bands (e.g. $m = \pm 1$ bands). To prove this, we fix the incident wave with a center frequency at 1kHz, which can resonantly excite the atom, then we measure the reflection amplitude spectrum by varying the modulation duty cycle $\xi$ from 0 to 1, as a 2D color map in fig. 2**c**. For $\xi = 0.5$, the reflection amplitude is almost zero for the main-band ($m = 0$), in contrary to the side bands at $m = \pm 1$ with significant reflection amplitudes, i.e. with the highest efficiency of the frequency conversion to the side-bands. We note that for $\xi = 0$ ($\xi = 1$), the atom goes back to the static atom with pure gain (loss), no frequency conversion occurs thus no side bands are obtained as expected. For further illustration that gain-loss balance diminishes the main-band, we have also performed another series of experiments that by choosing cases with $a_1 \neq a_2$ with variable duty cycle $\xi$ but keeping $\bar{a} = 0$ (see supplementary materials).

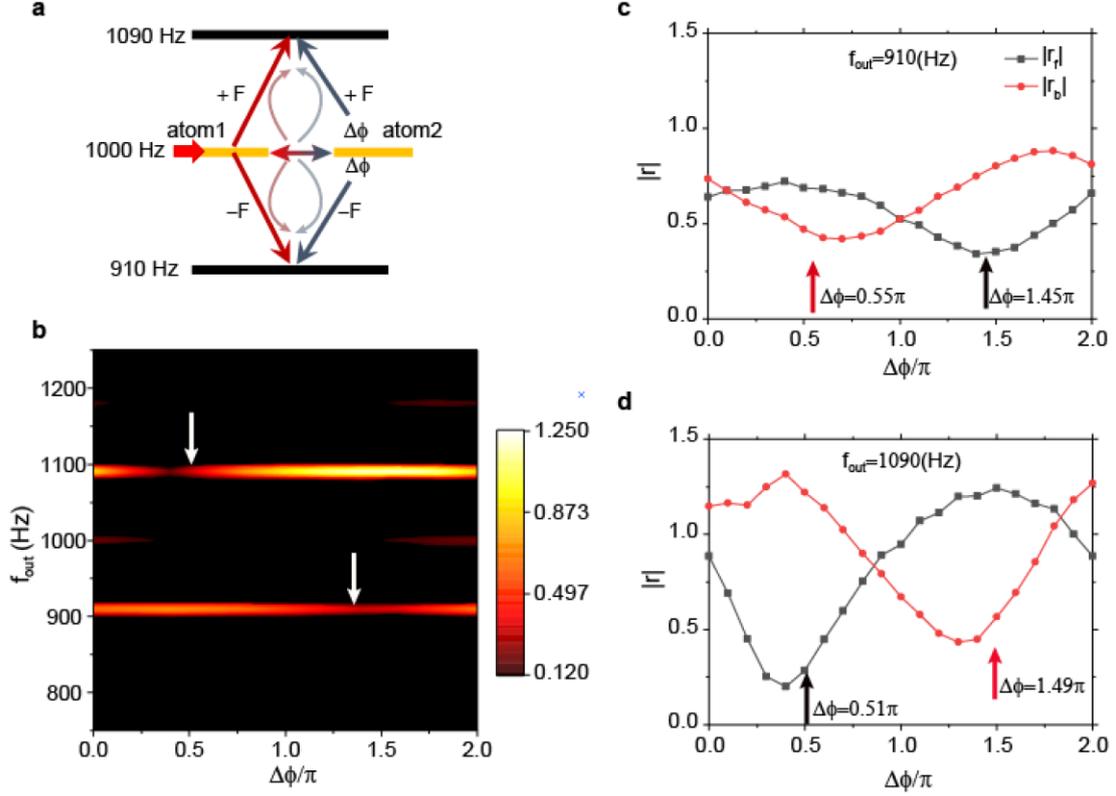

**Fig. 3| Spatially asymmetric frequency conversion and amplification. a**, The "energy level diagram" for two time-varying atoms with the same resonance frequency (the yellow bars) and modulation frequency $F$, and a modulation phase delay $\Delta\phi$ (induced by the additional delay time $\Delta t$) is applied on atom2. Here the thick arrow represents the incident frequency and thin arrows represent the directions of frequency conversion. **b**, The 2D map of the forward reflection amplitude spectrum ($|r_f|$) against the modulation phase delay $\Delta\phi$ and the output frequency. The input frequency is kept at 1kHz. The magnitudes of the side bands exhibit interference characteristics, where arrows indicate the destructive dip position calculated from the Floquet harmonic model. The small discrepancy in location is due to the electronic delay in the microcontroller. In addition, the forward and backward reflection amplitudes spectra with different interference conditions reveal the asymmetric scattering property for both $m=-1$ (**c**) and $m=1$ (**d**) side band.

Now, we go to the case with two atoms construct a space-time varying metamaterial to enable more control (through interference) on the frequency conversion. Figure 3**a** shows such a configuration for a pair of atoms with the same resonating frequency (1kHz) and modulation frequency (90Hz). With the previously verified transfer process, we choose to describe the configuration using an energy level diagram. The yellow levels describe those with resonance. The levels with black color denote those allowed levels satisfying frequency transition rule: $f_{out} = f_{in} + mF$. The thin arrows indicate the corresponding frequency transitions. Each arrow must start from a resonating level but can end at any. In the current case, we only focus on the $m = 1$ and $m = -1$ transitions by assuming an incident wave at 1kHz. Different from the case of a single atom, there can be more than one path to get a specific frequency conversion performed. The incident wave resonantly excites atom 1 and have a frequency transition to 910Hz by a 1$^{st}$ order down conversion ($m = -1$). On the other hand, the incident wave equivalently excites atom 2 and have the same frequency down conversion. These two paths can now undergo interference with each other, in a way mimicking quantum interference, which is the fundamental building block in many quantum phenomena, such as electromagnetically induced transparency [29]. Such an interference can in fact be affected by numerous factors, such as the distance between the two atoms, the program time-delay $\delta t$ in Eq. (1), and also the modulation phase delay (or spatial modulation) $\Delta\phi$ (or $2\pi F\Delta t$) between the two atoms. $\Delta\phi$ can vary from 0 (atoms in synchronization) to $2\pi$ by going through one cycle of modulation delay (see Fig. 1**b**). Fig. 3**b** shows the reflection amplitude spectrum when we vary $\Delta\phi$ for forward incidence (incident wave from the left in the waveguide). While the main band reflection remains low in magnitude from the gain-loss balance, we observe that the frequency conversion amplitude varies with $\Delta\phi$ on either the $m=1$ or $m=-1$ band, which contains significant conversion amplitude. Fig. 3c and d show the zoom-in version of the same conversion amplitude at the two bands (910 and 1090Hz in black (red) color for forward (backward) incidence). The dip locations of the amplitude results from a destructive interference between the two paths, take $m=1$ Floquet mode as an example, it is governed by

$$\pm F\Delta t + f_{in}\delta t_{21} + f_{out}\delta t_{12} = n + 1/2, n \in \mathbb{Z} \tag{2}$$

where $+(-)$ sign indicates for the case of forward (backward) incidence, and $\delta t_{ij}$ is the propagation time delay from microphone $D_i$ to speaker $S_j$. The results predicted from the interference conditions (see details in supplementary materials) are indicated by the four arrows in Fig. 3c,d, which agree reasonably well with experimental dips.

From the Eq. (2), $\Delta\phi$ introduces a spatially asymmetric control on the conversion efficiency. At output frequency 1090Hz and $\Delta\phi = 0.4\pi$, the strong reflection amplitude in the backward incidence has nearly 7 times the amplitude in the forward direction. We also emphasize that in one direction, the power output from the frequency conversion exceeds the incident power by nearly 30% at main band (with conversion efficiency 1.3). For the 1st order side-band ($f_{out} = 1090Hz$), the reflection amplitude can exceed 1, i.e. the efficiency of frequency conversion larger than 100%. This is because the secondary radiation source in the meta-atoms draw power from the external digital circuits, while other time-varying approaches are difficult to achieve such high efficiency up to our knowledge.

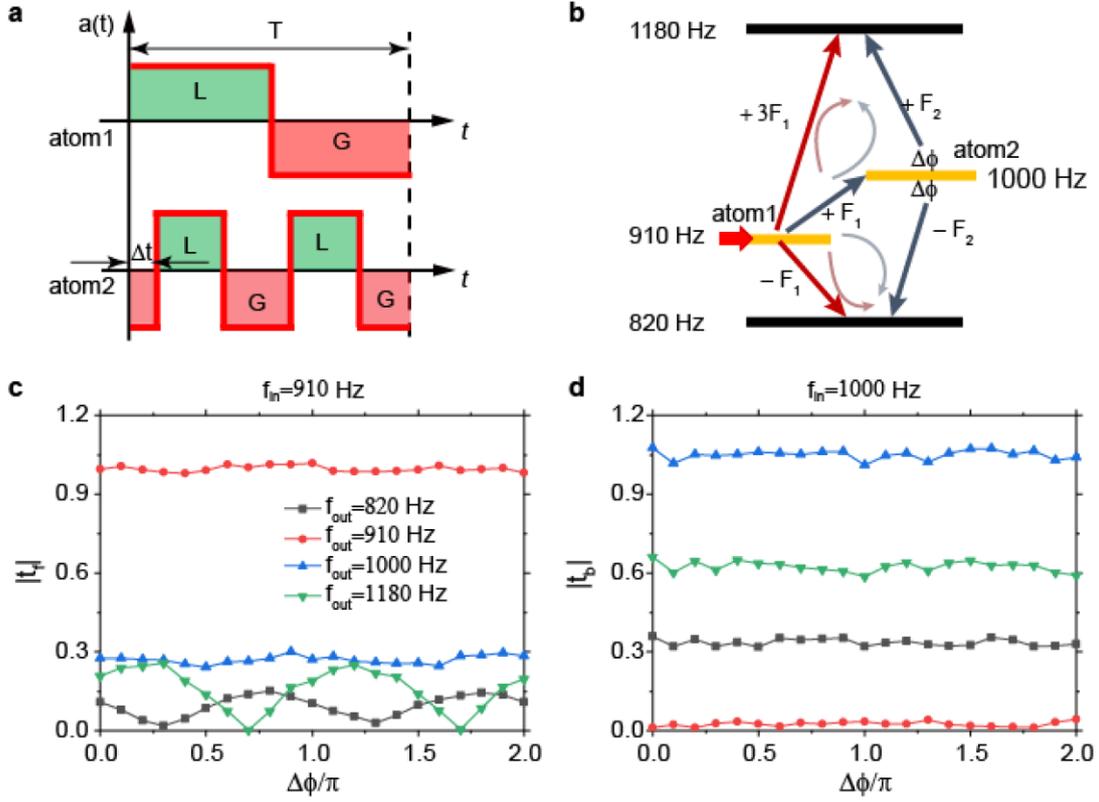

**Fig. 4|Unidirectional frequency coupling between two atoms with interference control. a**, The schematic of amplitude modulation for two time-varying atoms, where atom2 has doubled modulation frequency ($F_2 = 2F_1 = 180Hz$) and an additional modulation time delay $\Delta t$. **b**, The "energy level diagram" for two time-varying atoms with resonating frequency at 910Hz and 1000Hz respectively, where the modulation phase delay is defined as $\Delta\phi=2\pi F_1\Delta t$. For both output frequencies at 820Hz and 1180Hz, there are two paths to get to the same energy level. **c**, Forward transmission amplitude $|t_f|$ spectra at various output frequencies: 820Hz (black), 910Hz (red), 1000Hz (blue) and 1180Hz (green) for input frequency at 910Hz. **d**, Backward transmission amplitude $|t_b|$ at the similar output frequencies for input frequency at 1000Hz. Due to the unidirectional coupling from atom 1 to 2 (910 → 1000Hz) versus no coupling the other way round from atom 2 to 1 (1000 → 910Hz), the transmission amplitude differ as blue (red) line in fig. c (d).

As our metamaterial platform allows software-defined resonating response and modulation specification for each atom, the energy level diagram and the interference pathways for frequency conversion can be engineered freely. Figure **4b** shows a representative example that the modulation frequencies for the two atoms are now different, allowing a unidirectional coupling/frequency conversion between the two atoms. In the current case, we design atom1 (atom2) has a resonating frequency at 910Hz (1000Hz) with modulation frequency $F_1$ =90Hz ($F_2$ =180Hz). As long as there exists a common multiple of the two modulation periods $T_1$ and $T_2$, denoted as $T$ by $T = n_1 T_1 = n_2 T_2$ with integer $n_1$ and $n_2$, $\Delta\phi = 2\pi F_1 \Delta t$ can be meaningfully defined from 0 to $2\pi$ along with a shifting of $n_1$ and $n_2$ cycles across $T$ for the two modulations, see Fig. **4a** for the current case of $n_1 = 1$ and $n_2 = 2$ with $\Delta\phi$ defined on atom2. In addition, the energy diagram also constitutes a set of well-defined discrete levels with the existence of such a common multiple $T$, which is now shown in Fig. **4b**. The energy level diagram is tailor-made to allow multiple pathways and with unidirectional coupling feature. For an incident wave at 910Hz, resonating at atom1, a 1st order transition brings it to 1000Hz while the same incident wave does not interact much with atom2. On the other hand, for an incident wave at 1000Hz, resonating at atom2, a 1st order transition can only bring it to 1180Hz or 820 Hz with significant efficiency, leaving nearly no waves emitted at 910Hz. Such a unidirectional frequency coupling (between the two resonating levels of the two atoms) is demonstrated in the current configuration. The blue curve in Fig. **4c** shows a forward transmission amplitude around 0.3 for conversion 910Hz→1000Hz while the red curve in Fig. **4d** shows a much smaller backward transmission amplitude for conversion 1000Hz → 910Hz ($<0.05$). Such a transmission amplitude contrast between forward and backward directions in converting between two modes is the result of the above unidirectional coupling and shares similar features with nonreciprocal transmission [14, 24, 25]. In this case, the unidirectional frequency conversion is independent of $\Delta\phi$ as there is only at most one single pathway dominating the process. The asymmetry in transmission conversion amplitudes is due to the different designs of modulated resonances of the two atoms.

Finally, we go on to discuss the interference effect enabled by $\Delta\phi$ for the other two energy levels (820Hz and 1180Hz). For an incident frequency of 1000Hz in resonance with atom2, only single pathway is allowed by a first order transition to either frequency, as shown as the flat black and blue curve in Fig. 4**d**, i.e. interference is off. On the other hand, for an incident frequency of 910Hz in resonance with atom1, interference is turned on. The transmission amplitudes (black and green for the two frequencies) exhibit constructive and destructive interference peaks and dips when we vary $\Delta\phi$. Such a phenomenon can be explained using the energy level diagram. As an example for the frequency conversion from 910Hz to 820Hz, the incident wave resonates with atom1, down converts to 820Hz by an order 1 transition as one path. Alternatively, up conversion also happens to 1000Hz at atom1, further resonating at atom2 and a subsequent down conversion to 820Hz by an order 1 transition ($-180Hz$) at atom2 as another path. These two paths interfere with each other to create the dips and peaks in forward transmission amplitude spectrum (against $\Delta\phi$). We also note that there are now two dips (and two peaks as well) due to the fact that $n_2 = 2$ in our current example (see Supplementary Information for the interference conditions). For the frequency conversion process from 910Hz to 1180Hz, there are again two pathways: a 3$^{rd}$ order up-conversion at atom1 is one path while a 1$^{st}$ order up-conversion at atom1 to atom2 to 1000Hz undergoes a further 1$^{st}$ order up-conversion at atom2 to 1180Hz as another path. The dynamics of these multiple pathways are depicted with the arrows in Fig. 4**b**. The above has demonstrated that the "energy-level" diagrams can be freely engineered to have tailor-made allowed (or forbidden) transitions.

While the frequency conversion efficiencies in converting $f_1$ to $f_2$ is the same $f_2$ to $f_1$ (with $f_1 \neq f_2$) for a conventional system, such a rule can be largely broken when we consider a time-varying non-Hermitian system. Instead of using non-Hermitian coupling term that can be possible in a time-varying system [28], the current approach uses different modulation frequencies for the two atoms together with material gain to induce significantly different frequency conversion efficiencies between $f_1$ to $f_2$ and vice-versa. In other words, in a Hamiltonian picture in describing the dynamics, we are using the non-Hermiticity in the diagonal matrix elements directly instead in the off-diagonal ones. The current approach thus provides an alternative but straightforward way to achieve very asymmetric frequency conversion (comparing $f_1$ to $f_2$ in the forward and either $f_1$ to $f_2$ or $f_2$ to $f_1$ in the backward direction). Certainly, such a phenomenon is enabled by our flexible time-varying metamaterial platform in the current work while the ability of gain-loss balance in the time-domain makes the frequency conversion very efficient by diminishing the main band.

We emphasize that due to the active nature of our system, the frequency conversion is much more efficient than conventional approaches. The power required to the output channel is actually provided by the external source, i.e. the electric power injected into the microcontroller. Based on our current demonstrated examples, we believe our approach in realizing time-varying non-Hermitian systems have a lot of opportunities for further extension. Using the energy level diagrams as the language, our platform can certainly allow more atoms/levels for sophisticated pathways, e.g. more than 2 pathways in the current work. Moreover, more than one resonating levels are also allowed even for a single atom, which allows further exploration of time-varying and non-Hermitian physics based on the current platform.